\documentclass[10pt]{iopart}

\usepackage{epsfig}  
\begin{document}

\title[The Monte-Carlo dynamics of a binary Lennard-Jones 
glass-forming mixture]{The Monte-Carlo dynamics of a binary Lennard-Jones 
glass-forming mixture}

\author{L. Berthier and W. Kob}
\address{Laboratoire des Collo{\"\i}des, Verres
et Nanomat{\'e}riaux, UMR 5587, Universit{\'e} Montpellier II and CNRS,
34095 Montpellier, France}
\ead{berthier@lcvn.univ-montp2.fr, kob@lcvn.univ-montp2.fr}

\begin{abstract}
We use a standard Monte-Carlo algorithm 
to study the slow dynamics of a binary 
Lennard-Jones glass-forming mixture at low temperature.
We find that Monte-Carlo is by far the most
efficient way to simulate a stochastic 
dynamics since relaxation is about 10 times 
faster than in Brownian Dynamics and about 
30 times faster than in Stochastic Dynamics.  
Moreover, the average dynamical behaviour of the system 
is in quantitative agreement with the one obtained using 
Newtonian dynamics, 
apart at very short times where thermal vibrations are suppressed. 
We show, however, that dynamic fluctuations quantified 
by four-point dynamic susceptibilities do retain a dependence 
on the microscopic dynamics, as recently predicted theoretically. 
\end{abstract}


\section{Introduction}

Numerical simulations play a major role among 
studies of the glass transition
since, in contrast to experiments,
the individual motion of a large number of particles 
can be followed at all times~\cite{hans}. 
With present day computers, it is possible to follow the dynamics 
of a simple glass-forming
liquid over more than 8 decades of time, and over a temperature 
window in which average relaxation timescales increase by
more than 5 decades. However, 
at the lowest temperatures studied, relaxation is still
orders of magnitude faster than in experiments performed 
close to the glass transition temperature. Nevertheless,  
it is now possible to numerically access temperatures which are low 
enough that many features associated to the glass transition physics
can be observed: Strong decoupling phenomena~\cite{harrowell2,onuki,berthier}, 
clear deviations from fits to the mode-coupling theory~\cite{KA} (which are
experimentally known 
to hold only at high temperatures), and 
crossovers towards activated dynamics~\cite{I,II}.

Computer simulations usually study Newtonian Dynamics (ND) by solving 
a discretized version of Newton's equations
for a given pair interaction  between particles~\cite{at}. Here,
we study a glass-forming model in which a binary mixture 
of small and large particles interact via a Lennard-Jones pair potential, 
a model introduced by Kob and Andersen (KA)~\cite{KA}.  
It can be interesting to study also different types of 
microscopic dynamics for the same pair potential. 
If dynamics satisfies detailed balance with respect to 
the Boltzmann distribution, all structural quantities remain unchanged, 
although the dynamics might be very different.
In colloidal glasses, for instance, 
the particles undergo Brownian motion arising
from collisions with the molecules of the solvent, and a stochastic 
dynamics is more appropriate. Theoretical considerations 
also suggest the study of different dynamics. Gleim et al. studied a
Stochastic Dynamics (SD) to investigate whether the relaxation
of the KA binary mixture depend on its microscopic dynamics, their answer
being ``no''~\cite{gleim}. In SD, a friction term and a random noise 
are added to Newton's equations, the amplitude of both terms 
being related by a fluctuation-dissipation theorem.
Szamel and Flenner recently used Brownian
Dynamics (BD) to study the same KA mixture~\cite{szamel2}. In this description 
there are no momenta, and positions evolve with a Langevin dynamics. 
They again find that relaxation using BD is very similar to the one 
resulting from ND. They emphasize that even the deviations 
from mode-coupling fitting are similar in BD and ND and conclude
that momenta play no role in avoiding the mode-coupling 
singularity, contrary to previous claims~\cite{previous}, 
but in agreement with more recent ones~\cite{more}.

Recently, it was also  discovered that dynamic heterogeneity, that is, 
spatio-temporal fluctuations around the average dynamical behaviour, 
sensitively depends upon the microscopic dynamics~\cite{I,II,science}. 
In particular, 
a major role is played by conservation laws 
for energy and density. In the case
of energy the mechanism can be physically understood as follows. For a
rearrangement to take place in the liquid, the system has to locally 
cross an energy barrier. If the dynamics conserves the energy, 
particles involved in the rearrangment must borrow energy to the 
neighboring particles. This `cooperativity' might be unnecessary if
energy can be locally supplied to the particles by an external heat bath.
Conservation laws, therefore, might 
introduce dynamic correlations between particles and dynamic
fluctuations can be different when changing from Newtonian
energy conserving dynamics to a stochastic thermostatted dynamics.
This predicted influence of the microscopic dynamics on dynamic
fluctuations~\cite{I,II} was in fact our principal motivation for 
the present study.

In this article, we propose a third type of stochastic dynamics
for the KA mixture and study in detail the dynamics of the system 
subjected to a standard Monte-Carlo (MC) dynamics.  We find that
MC is particularly efficient at relaxing the system since it is 
about 10 times faster than BD and 30 times faster than SD, 
while the average dynamics is still in quantitative agreement with ND.
We are therefore in position to study both the very low 
temperature average 
dynamics of the model and its dynamic fluctuations in detail, 
shedding new light on both aspects.

The paper is organized as follows. In Section~\ref{mc} we give 
details about the simulation technique and compare its efficiency 
to previously studied dynamics. In Section \ref{results} we present 
our numerical results. Section \ref{conclusion} concludes 
the paper.  

\section{An efficient simulation technique}
\label{mc}

We study a binary Lennard-Jones mixture made of 
$N_A=800$ and $N_B=200$ 
particles of types $A$ and $B$, respectively.  
Particles interact with the following Lennard-Jones pair potential
\begin{equation} 
\phi_{\alpha \beta}^{\rm LJ}(r)= 
4 \epsilon_{\alpha \beta} \left[ 
\left(\frac{\sigma_{\alpha \beta}} {r} 
\right)^{12} - 
\left(\frac{\sigma_{\alpha \beta}}{r} 
\right)^{6} 
\right], 
\label{pp}
\end{equation} 
where $\alpha, \beta \in [{A}, {B}]$ and 
$r$ is the distance between the interacting pair of particles.
Interaction parameters $\epsilon_{\alpha \beta}$
and $\sigma_{\alpha \beta}$
are chosen to prevent crystallization
and can be found in Ref.~\cite{KA}. 
The length and energy are given in the standard 
Lennard-Jones units $\sigma_{AA}$ (particle diameter), and
$\epsilon_{AA}$ (interaction energy), where
the subscript $A$ refers to the majority  species.
The potential is truncated and shifted at a distance $r = 2.5$. 
Previous work~\cite{hans,KA} has shown that the dynamics becomes slow below 
$T \approx 1.0$, while the fitted mode-coupling temperature for this
system is $T_c \approx 0.435$, although deviations
from mode-coupling behaviour become 
noticable already below $T \approx 0.47$.   

We have implemented a standard Monte-Carlo dynamics~\cite{at} 
for the pair potential in Eq.~(\ref{pp}).  
An elementary move can be described as follows. 
A particle, $i$, located at the position ${\bf r}_i$  
is chosen at random. The energy cost, $\Delta E_i$, to move 
particle $i$ from position ${\bf r}_i$ to a new 
position ${\bf r}_i + \delta {\bf r}$ is evaluated,   
$\delta {\bf r}$ being a random vector comprised in a cube 
of linear length $\delta_{\rm max}$ centered around the origin.
The Metropolis acceptance rate, $p = {\rm min} 
(1, e^{-\beta \Delta E_i})$, where $\beta =1/T$ is the inverse temperature,
is then  used to decide whether the move is accepted.  
In the following, one Monte-Carlo timestep represents 
$N=N_A + N_B$ attempts to make such an elementary move, and timescales
are reported in this unit.

\begin{figure}
\begin{center}
\psfig{file=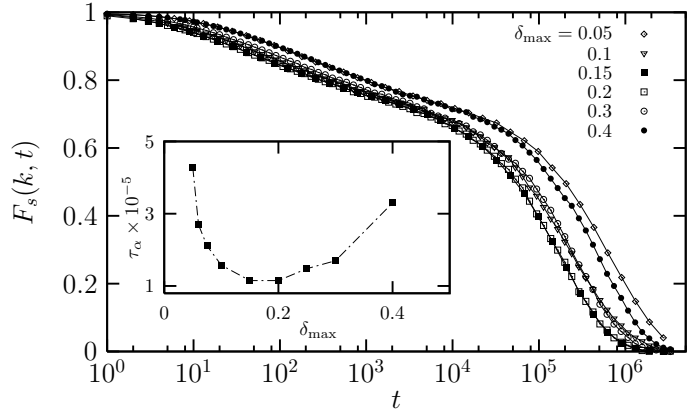,width=9.cm}
\end{center}
\caption{\label{delta} Self-intermediate scattering function,
Eq.~(\ref{self}), at $T=0.5$ and $k=7.21$ for various values of 
$\delta_{\rm max}$.
Inset: The evolution of the relaxation time with 
$\delta_{\rm max}$ unambiguously defines an optimal value
$\delta_{\rm max} \approx 0.15$ for our simulations.}
\end{figure}

The one degree of freedom that remains to be fixed is $\delta_{\rm max}$
which determines the average lengthscale of elementary moves. 
If chosen too small, energy costs are very small and most of the moves
are accepted, but the dynamics is very slow because it takes a long 
time for particles to explore their cage.
On the other hand too large displacements will on average be very costly 
in energy
and acceptance rates can become prohibitively small. We seek a compromise 
between these two extremes by monitoring the 
dynamics at a moderately low temperature, $T=0.5$, for several  
values of $\delta_{\rm max}$. 
As the most sensitive indicator of the relaxational behaviour we measure the 
contribution from the majority specie $A$ to the 
self-intermediate scattering function, 
\begin{equation} F_s({\bf k},t) = 
\left\langle \frac{1}{N_A} \sum_{j=1}^{N_A} e^{i {\bf k} 
\cdot [{\bf r}_j(t) - {\bf r}_j(0)]} \right\rangle. 
\label{self} 
\end{equation}
We spherically average over wavectors of comparable magnitude, 
and present results for $|{\bf k}|=7.21$, which corresponds to the first
diffraction peak in the static structure factor of the liquid.
In Fig.~\ref{delta} we present our results for $\delta_{\rm max}$ 
values between 0.05 and 0.4. 
As expected we find that relaxation is slow both at small and large 
values of $\delta_{\rm max}$, and most efficient for intermediate values.  
Interestingly we also note that the overall 
shape of the self-intermediate
scattering function does not sensitively depend 
on $\delta_{\max}$. 

We define a typical relaxation time as 
$F_s(k,\tau_\alpha) = e^{-1}$ and show its $\delta_{\rm max}$ dependence
in the inset of Fig.~\ref{delta}. A clear minimum is observed
at the optimal value of $\delta_{\max} \approx 0.15$. In the rest
of the paper we only present data obtained for this value. 

As compared to previously studied dynamics,
we find that, when expressed in 
numbers of integration timesteps, structural relaxation in Monte-Carlo
simulations is marginally faster 
than in Newtonian dynamics, but 
30 times faster than in Stochastic Dynamics~\cite{gleim}, 
and 10 times faster 
than in Brownian Dynamics~\cite{szamel2}. We conclude therefore 
that MC is by far the most efficient way to perform 
stochastic molecular simulations of the present glass-forming material. 

The relative inefficiency of both BD and SD is due to the 
stochastic nature of their microscopic 
equations of motion. It is well-known
that small integration timesteps are required for accurate 
integration of stochastic equations of motion, in particular
to maintain the delicate balance between friction and noise 
required for the system to converge towards the 
correct equilibrium distribution~\cite{at}. No such constraint exists for 
MC dynamics, where elementary moves can be made arbitrarily 
large. Equilibrium only requires detailed balance to be fulfilled, and
this is always the case with the Metropolis algorithm described above.  
With larger elementary moves, particles can 
efficiently explore their cage 
and relaxation is much faster.
This physical interpretation is also supported by the optimal value 
$\delta_{\max}=0.15$ that we report, which corresponds to a 
mean-square displacement of 0.225, very close to the plateau
observed in the mean-square displacement shown
in Fig.~\ref{fs} (see below), which can be taken as a rough
estimate of the cage size. 
Monte-Carlo simulations can of course be made even more efficient by 
implementing for instance swaps between particles, or using 
parallel tempering. The dynamical behaviour, however, is then strongly
affected by such non-physical moves
and only equilibrium thermodynamics can be studied. Since 
we want to conserve 
a physically realistic dynamics, we cannot use such improved schemes.

We have performed simulations at temperatures between 
$T=2.0$ and $T=0.43$, the latter being smaller than
the fitted mode-coupling temperature.
For each temperature we have simulated 10 independent samples 
to improve the statistics. Initial configurations 
were taken as the final configurations obtained from 
previous work performed with ND~\cite{I,II}, so that production 
runs could be started immediately. For each sample, production 
runs lasted at least $15\tau_\alpha$ (at $T=0.43$),
much longer for higher temperatures. 

\section{Results}
\label{results}

\subsection{Average dynamics}

\begin{figure}
\hspace*{-0.2cm}
\psfig{file=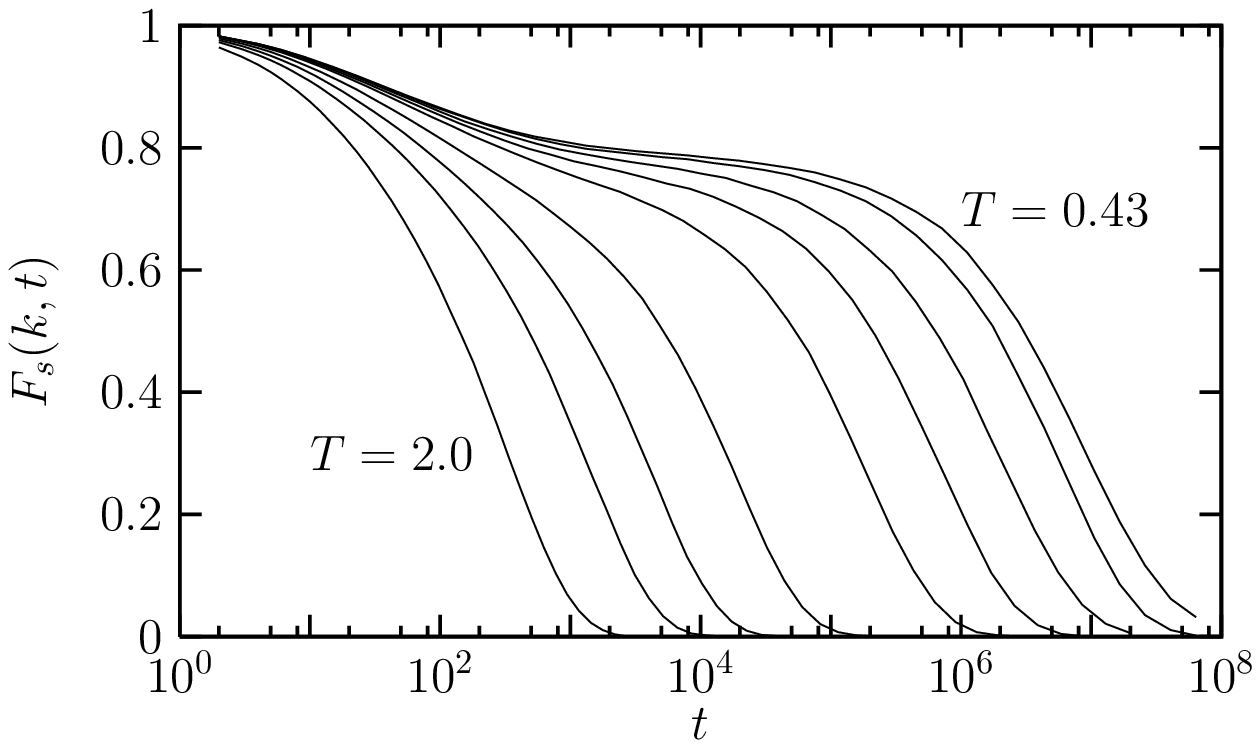,width=7.1cm}
\hspace*{-0.2cm}
\psfig{file=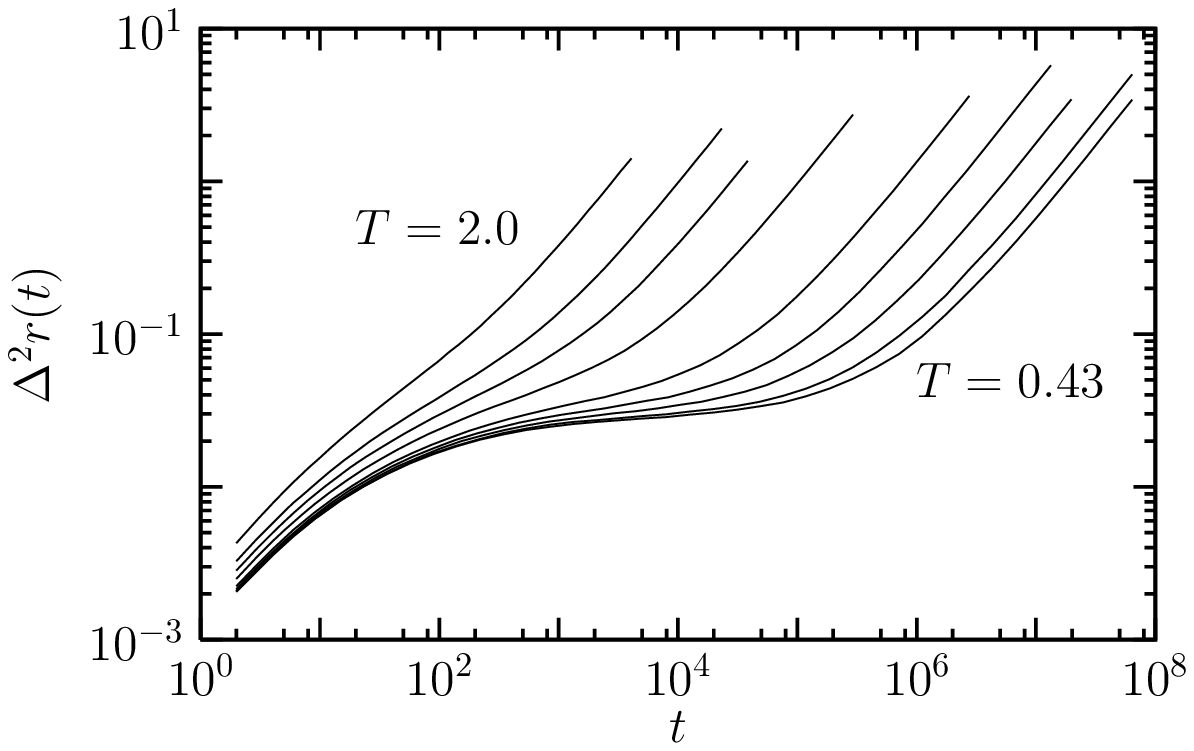,width=6.9cm}
\caption{\label{fs} Left: Self-intermediate scattering function, 
Eq.~(\ref{fs}), for $k=7.21$ and  
temperatures $T=2.0$, 1.0, 0.75, 0.6, 0.5, 0.47, 0.45, 0.435, and
0.43 (from left to right).
Right: Mean-squared displacement, Eq.~(\ref{msd}), for the same 
temperatures in the same order.}
\end{figure}

The self-intermediate scattering function, Eq.~(\ref{self}), 
is shown in Fig.~\ref{fs}
for temperatures decreasing from $T=2.0$ down to $T=0.43$. 
These curves present well-known features. Dynamics at high temperature 
is fast and has an exponential nature. When temperature is decreased
below $T \approx 1.0$, a two-step decay, the slower being strongly 
non-exponential, becomes apparent. Upon decreasing 
the temperature further,
the slow process dramatically slows down by about 5 decades, 
while clearly conserving an almost temperature-independent 
non-exponential shape, as already reported for ND~\cite{KA}.

Finally, as reported 
for SD~\cite{gleim}, we find that also the first process, the decay towards
a plateau, slows down considerably when decreasing temperature. 
This process, called `critical decay' in the language 
of mode-coupling theory~\cite{mct}, is not observed when using 
ND, because it is obscured by the thermal vibrations
occuring at high frequencies. 
Although the plateau seen in $F_s(k,t)$ is commonly
interpreted as `vibrations of a particle within a cage',
the data in Fig.~\ref{fs} discard this view. From direct
visualisation of the particles' individual dynamics
it is obvious that vibrations take place in just a few MC
timesteps, while the decay towards the plateau can be 
as long as $10^4$ time units at the lowest temperatures studied here. 
This decay is therefore necessarily more complex, 
most probably cooperative in nature. This interpretation
is supported by recent theoretical studies where a plateau 
is observed in two-time correlators of lattice models where local
vibrations are indeed completely absent~\cite{bethe}. 
A detailed atomistic description 
of this process has not yet been reported, but would 
indeed be very interesting.

Next, we study the mean-squared displacement for the majority
specie. It is defined as 
\begin{equation}
\label{msd}
\Delta^2 r(t) = \frac{1}{N_A} \sum_{i=1}^{N_A} \left\langle 
 |{\bf r}_i(t) - {\bf r}_i(0) |^2 \right\rangle,
\end{equation}
and we present its temperature evolution in Fig.~\ref{fs}, which
mirrors the evolution of the self-intermediate scattering function
in the same figure. Since we are studying a stochastic dynamics, 
displacements are diffusive at both short and long timescales. 
The plateau observed in $F_s(k,t)$ now translates into a sub-diffusive
regime in the mean-squared displacements separating the two diffusive
regimes. At the lowest temperature studied, when $t$ changes by three decades
from $2\times10^2$ to $2\times 10^5$, the mean-squared displacement
changes by a mere factor 2.2 from 0.02 to 0.044.
Particles are therefore nearly arrested for several decades of times,
before eventually entering the diffusing regime which 
allows for the relaxation of the structure of the liquid.  

\subsection{Comparison to Newtonian and Stochastic Dynamics}

The previous subsection has shown that the Monte-Carlo dynamics
of the KA mixture is qualitatively similar to the one 
reported for ND,
apart at relatively short times where the effect of thermal vibrations 
is efficiently suppressed. We now compare our results more
quantitatively with the dynamical behaviour observed using ND. 
 
\begin{figure}
\hspace*{-0.2cm}
\psfig{file=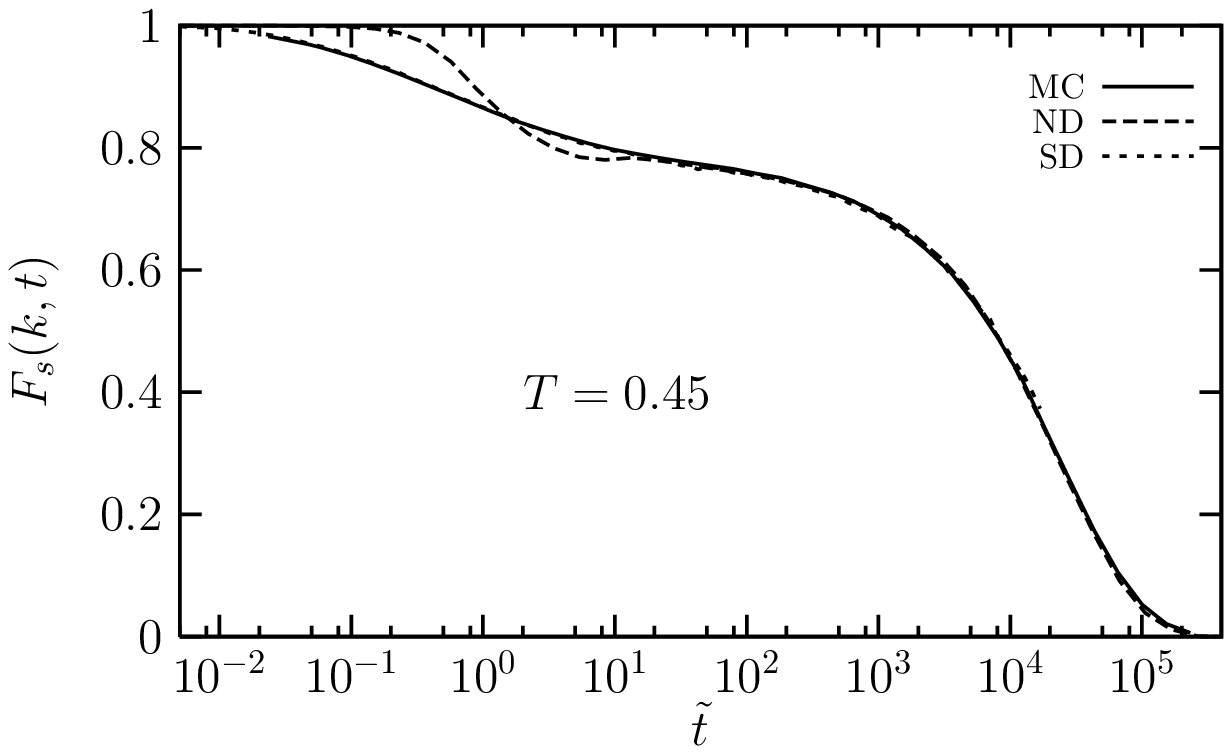,width=7.1cm}
\psfig{file=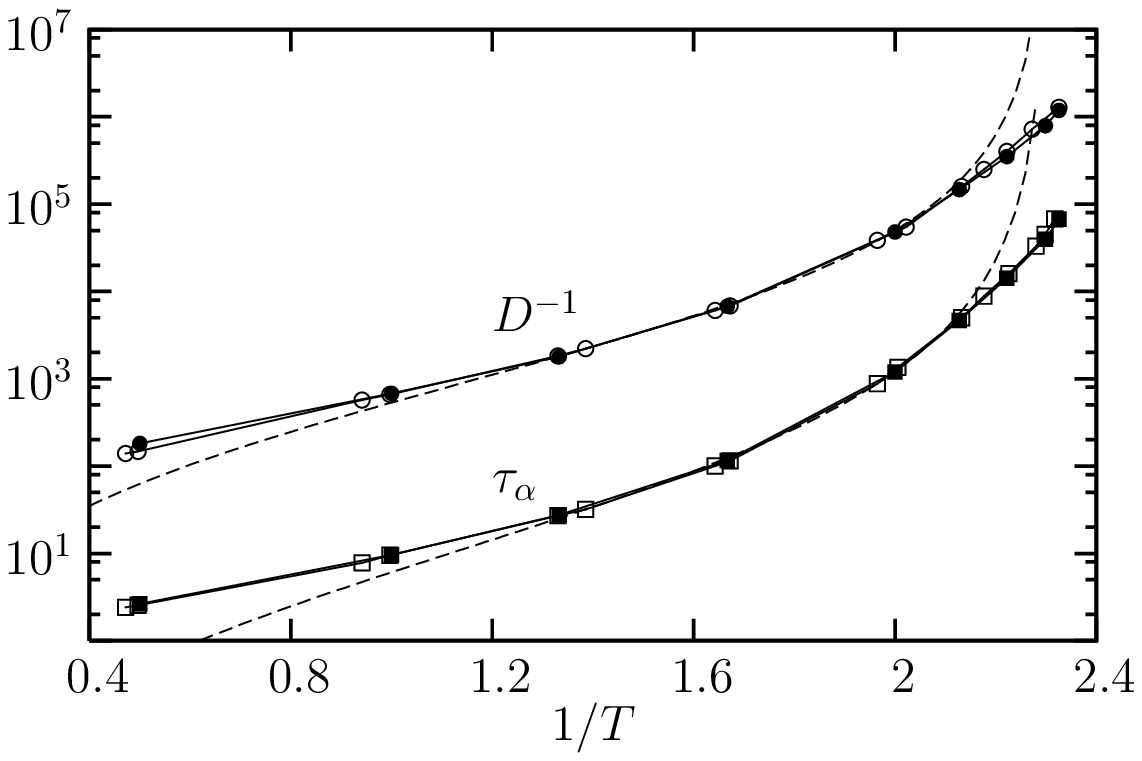,width=6.7cm}
\caption{\label{comp} 
Left: Comparison of the self-intermediate scattering 
function for $k=7.21$, and $T=0.45$, obtained 
in Monte-Carlo (MC) dynamics in this work, 
Newtonian Dynamics (ND) in Ref.~\cite{berthier}, 
and Stochastic Dynamics (SD) in Ref.~\cite{gleim}.
Time is rescaled to obtain maximum overlap at large times.
MC and SD agree over the complete time range (indeed the SD dotted
line is barely visible below the full MC line), while 
MC and ND only agree when $F_s(k,t)$ is close to the plateau
and below. The dip observed at short-time in ND is due to thermal
vibrations suppressed in both SD and MC.
Right: Temperature evolution of the alpha-relaxation time
$\tau_\alpha(T)$ and the inverse of the self-diffusion constant $1/D(T)$
in an Arrhenius plot. Open symbols are for ND, closed symbols for MC
(vertically shifted to obtain maximum overlap with ND data) 
the dashed lines are power law fits to a divergence at $T_c=0.435$, 
as originally reported in Ref.~\cite{KA}.}
\end{figure}

In Fig.~\ref{comp} we compare the time dependence 
of the self-intermediate scattering function for three types of dynamics:
the present Monte-Carlo data, the Newtonian Dynamics data taken from 
Ref.~\cite{I}, and the 
Stochastic Dynamics results from Ref.~\cite{gleim}, all obtained for 
the same parameters $k=7.21$ and $T=0.45$.  
We have rescaled the time to obtain 
maximum overlap in the long-time relaxation
of the three curves. Quite strikingly, SD and MC data perfectly overlap 
over the complete time-range (8 decades of time) 
of the simulation. Indeed the SD dotted 
line is barely visible below the full line of the MC data 
in Fig.~\ref{comp}. This confirms our claim that MC defines 
a physically relevant microscopic dynamics, since it is 
completely equivalent to SD with the major 
advantage that it is 30 times faster, at least for the KA mixture.

In Fig.~\ref{comp}, we also confirm that the approach 
to the plateau is different in MC/SD and ND. In the latter, phonon-like
vibrations affect the initial decay of $F_s(k,t)$. For instance, 
a shallow dip, generally attributed to the `Boson peak', is observed
at low temperature in ND, see the dashed line in Fig.~\ref{comp}. 
The long-time decay of the self-intermediate scattering function, however,
is in full quantitative agreement for the three dynamics. This 
agreement was the main
claim of Ref.~\cite{gleim}, extended to BD in Ref.~\cite{szamel2}
and for MC in the present work.

Since all dynamics display similar long-time relaxation, it is sensible
to also quantitatively compare the temperature evolution of the 
relaxation times, $\tau_\alpha(T)$, already defined above.
This is done in Fig.~\ref{comp}, where we use a standard 
representation where an Arrhenius slowing down over a 
constant energy barrier, $\tau_\alpha \sim \exp( E/T)$, would 
appear as a straight line. The data clearly show some upwards 
bending in Fig.~\ref{comp}, which places the KA mixture in the family
of fragile (though very weakly) glass-formers.
We find that the temperature 
evolution of the alpha-relaxation time measured in MC simulations
is in complete quantitative agreement with the 
one obtained from ND, over the complete temperature range
$T=2.0 \to 0.43$. In particular the quality 
of a power-law fit of the slowing down, $\tau_\alpha 
\sim (T-T_c)^{-\gamma}$,
as suggested by mode-coupling theory, is similar for both 
dynamics~\cite{KA,gleim}. 
We have shown such a fit through our data, using the value $T_c=0.435$
determined in Ref.~\cite{KA}. The fit describes the data over about 2.5 
decades. Deviations from the mode-coupling fit appear below
$T \approx 0.47$, and become obvious when $T_c$ is approached further.  

In Fig.~\ref{comp} we also show the temperature evolution
of the self-diffusion constant, defined from the long-time limit
of the mean-square displacement as 
\begin{equation}
D = \lim_{t \to \infty} \frac{\Delta^2 r(t)}{6 t}.
\end{equation}
The behaviour of the diffusion constant is qualitatively very close 
to the one of the alpha-relaxation time, and all the above remarks 
apply. The well-known difference between the two quantities 
is a slightly stronger temperature evolution of $\tau_\alpha$,
implying a well-studied decoupling between translational 
diffusion and structural relaxation in this system~\cite{hans,berthier}, 
which is therefore very similar for different types of dynamics.

Theoretically, an identical relaxation within MC/SD/BD/ND 
is an important prediction of mode-coupling theory~\cite{mct} because 
the theory uniquely predicts the dynamical behaviour from static
density fluctuations. Gleim et al. argue that their finding 
of a quantitative agreement between SD and ND is a nice confirmation
of this non-trivial mode-coupling prediction~\cite{gleim}. Szamel and
Flenner~\cite{szamel2} 
confirmed this claim using BD, and argued further that 
even deviations from mode-coupling predictions are identical.
We confirm the validity of this statement even below $T_c$, 
showing that the agreement between different dynamics, 
although indeed predicted by mode-coupling
theory, is certainly valid at a much more general level. 
Similarly to Szamel and Flenner, we note that deviations from a
power law divergence cannot be attributed to coupling to currents
which are expressed in terms of particle velocities.
In our MC simulations we have no velocities,
so that avoiding the mode-coupling singularity is not due to 
the hydrodynamic effects pointed out in Ref.~\cite{previous} 
(see Ref.~\cite{more} for more recent theoretical 
viewpoints).

\subsection{Multi-point susceptibility}

Having established the ability of MC simulations
to efficiently reproduce the average 
slow dynamics obtained from ND simulations we now turn
to the study of the dynamic fluctuations around the average dynamical
behaviour, i.e. to dynamic heterogeneity.

\begin{figure}
\begin{center}
\psfig{file=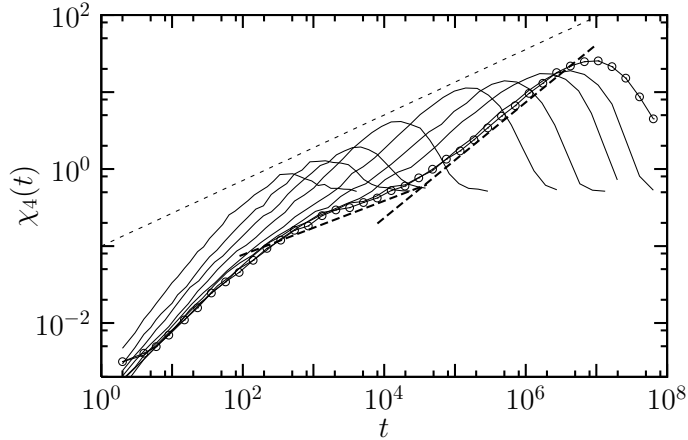,width=9.cm}
\end{center}
\caption{\label{chi4} Four-point susceptibility, Eq.~(\ref{chi4lj}),
for the same temperatures as in Fig.~\ref{fs}, decreasing from 
left to right. We have highlighted with open circles the data
at $T=0.43$, which are fitted with two power laws shown as 
dashed lines with exponents $0.35$ and $0.75$ at short and large times,
respectively.}
\end{figure}

Dynamic fluctuations can be studied through the  
four-point susceptibility, $\chi_4(t)$, which 
quantifies the strength of the spontaneous fluctuations 
around the average dynamics by their variance, 
\begin{equation} 
\chi_4(t) = N_A \left[  \langle f_s^2({\bf k}, t) 
\rangle - F_s^2({\bf k}, t) \right], 
\label{chi4lj} 
\end{equation} 
where $f_s({\bf k},t) = N_A^{-1} \sum_j  \cos ({\bf k} 
\cdot [{\bf r}_j(t) - {\bf r}_j(0)] )$ represents
the real part of the instantaneous value of 
the self-intermediate scattering function, 
so that $F_s({\bf k},t) = \langle f_s({\bf k},t) \rangle$.
As shown by Eq.~(\ref{chi4lj}), it is clear that 
$\chi_4(t)$ will be large if run-to-run fluctuations
of the self-intermediate scattering functions are large. This is
the case when the local dynamics becomes
spatially correlated, as already discussed in several 
papers~\cite{FP,silvio2,glotzer,lacevic,toni,mayer}.

We show the time dependence of the dynamic susceptibility
$\chi_4(t)$ obtained from our MC simulations for various temperatures 
in Fig.~\ref{chi4}. As predicted theoretically 
in Ref.~\cite{toni} we find that $\chi_4(t)$ presents
a complex time evolution, closely related to 
the time evolution of the self-intermediate scattering function.
Overall, $\chi_4(t)$ is small at both small and large 
times when dynamic fluctuations are small. There is therefore 
a clear maximum observed for times comparable to 
$\tau_\alpha$, where fluctuations are most prominent. The position
of the maximum then shifts to larger times when 
temperature is decreased, tracking the alpha-relaxation.
The most important physical information revealed by these curves is the
fact that the amplitude of the peak grows when the temperature 
decreases. This is direct evidence that spatial correlations grow 
when the glass transition is approached. 

The two-step decay of the self-intermediate scattering function
translates into a two-power law regime for $\chi_4(t)$ 
approaching its maximum. We have fitted these power laws,
$\chi_4(t) \sim t^a$, followed by $\chi_4(t) \sim t^b$ with the 
exponents $a =0.35$ and $b=0.75$ in Fig.~\ref{chi4}. We 
have intentionally used the notation $a$ and $b$ for these 
exponents which are predicted, within mode-coupling theory, 
to be equal to the standard exponents also describing
the time dependence of intermediate scattering functions~\cite{mct}. 
Our findings are in good agreement with previously 
reported values for $a$ and $b$. See Refs.~\cite{II,toni} for a more extensive
discussion and comparison to other theoretical predictions.

\begin{figure}
\begin{center}
\psfig{file=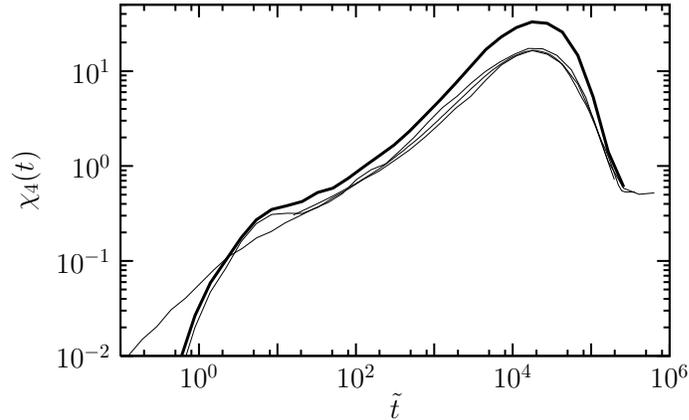,width=9.cm}
\end{center}
\caption{\label{chi42} Four-point susceptibility for various dynamics
and ensembles at $T=0.45$. As in Fig.~\ref{comp}, times
have been rescaled to obtain the maximum overlap in the 
long-time regime. 
The three overlapping thin lines represent data for  
ND-$NVE$ dynamics, SD, and MC, while the thick line represents
ND-$NVT$ data, for which dynamic fluctuations are clearly 
larger, as predicted theoretically and discussed in Ref.~\cite{I}.}
\end{figure}

We finally compare 
the dynamic susceptibility for various dynamics.
In Fig.~\ref{chi42}, we present the time evolution 
of $\chi_4(t)$ for a given temperature, $T=0.45$ and four 
different dynamics: 
The present MC data, data from SD obtained in Ref.~\cite{I}, 
data for ND in the microcanonical ($NVE$) ensemble from Ref.~\cite{I},
and data for ND in the canonical ($NVT$) ensemble from 
Ref.~\cite{I}. To perform this comparison, we have again rescaled times
to obtain the maximum overlap in the long-time region.
In Fig.~\ref{chi42} it is obvious that three curves are identical:
ND-$NVE$, MC and SD data perfectly overlap near the maximum of 
$\chi_4(t)$ and have similar time dependences, apart at very short-times.
On the other hand, ND-$NVT$ data display a different time dependence
and reveal considerably larger dynamic fluctuations in the long-time regime.
 
We conclude therefore that, contrary to the average dynamics, 
the dynamic fluctuations quantified through the four-point susceptibility
do retain a dependence upon the microscopic dynamics since canonical  
estimates of $\chi_4(t)$ are different for ND and for MC/SD/BD.
Although perhaps counter-intuitive at first sight we find that 
dynamics with a stochastic heat-bath display dynamic fluctuations
similar to the ones measured using microcanonical ND, while
fluctuations are much larger in canonical ND simulations. 
As mentioned in the introduction, this confirms the idea that
the energy conservation (implied by 
Newton's equations of motion) might lead to 
an amplification of dynamic fluctuations. With hindsight, this is 
not such a surprising result: The specific heat, after all,
also behaves differently in different statistical ensembles.
The ensemble dependence and dependence upon the microscopic dynamics
are the main subjects of two recent papers~\cite{I,II}.

There is
an experimentally relevant consequence of these findings. The difference
between the microcanonical and canonical values of the dynamic 
fluctuations in ND can be shown to be equal to~\cite{science}
\begin{equation}
\chi_4^{NVT}(t) - \chi_4^{NVE}(t) = \frac{T^2}{c_V} \left(
\frac{\partial F_s({\bf k},t)}{\partial T} \right)^2,
\label{chiT}
\end{equation}
where $c_V$ is the constant volume specific heat expressed in 
$k_B$ units. As shown in 
Fig.~\ref{chi42} the temperature derivative in Eq.~(\ref{chiT})
represents in fact the major
contribution to $\chi_4^{NVT}$, meaning
that the term $\chi_4^{NVE}$ can be neglected in Eq.~(\ref{chiT}).
Since the right hand side of (\ref{chiT}) is 
more easily accessible in an experiment than $\chi_4$ itself, 
Eq.~(\ref{chiT}) opens the possibility of an experimental
estimate of the four-point susceptibility. This finding, 
and its experimental application to supercooled glycerol and 
hard sphere colloids, 
constitute the central results of Ref.~\cite{science}.
   
\section{Conclusion}
\label{conclusion}

We have implemented a standard Monte-Carlo dynamics on the well-known
binary Lennard-Jones mixture introduced by KA.
We have shown that the resulting average 
dynamics is in full quantitative agreement with results from 
Newtonian dynamics, while being considerably faster
than previously studied stochastic dynamics, namely 
Brownian and Stochastic 
dynamics. We have therefore at our disposal 
an efficient numerical technique
to simulate the stochastic 
dynamics of the KA mixture at low temperature.
This allowed us to show, in particular, that dynamic fluctuations
retain a dependence upon the microscopic dynamics
since four-point dynamical susceptibilities 
evaluated in the canonical ensemble for ND and MC quantitatively
differ, because the energy conservation of Newton's equations
amplify dynamic fluctuations.

\ack We wish to thank J.L. Barrat for useful discussions, and 
G. Biroli, J.P. Bouchaud, K. Miyazaki, and D. Reichman 
for our recent collaboration~\cite{I,II}, which initially motivated 
this work.


\section*{References}

\begin{thebibliography}{10}


\bibitem{hans}  
H.~C. Andersen,  Proc. Natl. Acad. Sci. 
{\bf 102}, 6686 (2005). 

\bibitem{harrowell2} 
D.~N. Perera and P. Harrowell, J. Chem. Phys. {\bf 111}, 5441 (1999). 

\bibitem{onuki}
R. Yamamoto and A. Onuki, Phys. Rev. Lett. {\bf 81}, 4915 (1998).

\bibitem{berthier} L. Berthier, 
Phys. Rev. E {\bf  69}, 020201 (2004). 

\bibitem{KA} W. Kob and H.~C. Andersen, 
Phys. Rev. Lett. {\bf 73}, 1376 (1994); 
Phys. Rev. E {\bf 53}, 4134 (1995); Phys. Rev. E {\bf 51}, 4626 (1995). 

\bibitem{I} L. Berthier, G. Biroli, J.-P. Bouchaud,  
W. Kob, K. Miyazaki, D.~R. Reichman, cond-mat/0609656.

\bibitem{II} 
L. Berthier, G. Biroli, J.-P. Bouchaud,  
W. Kob, K. Miyazaki, D.~R. Reichman, cond-mat/0609658.

\bibitem{at} M. Allen and D. Tildesley,  
{\it Computer Simulation of Liquids} (Oxford University Press, Oxford, 1987). 

\bibitem{gleim} T. Gleim, W. Kob, and K. Binder, 
 Phys. Rev. Lett. {\bf 81}, 004404 (1998).  

\bibitem{szamel2} 
G. Szamel, and E. Flenner,
Europhys. Lett. {\bf 67}, 779 (2004). 

\bibitem{previous}  S.P. Das and G.F. Mazenko, Phys. Rev. A {\bf 34}, 2265 
(1986).

\bibitem{more}  A. Andreanov, G. Biroli and A. Lef\`evre,
J. Stat. Mech. P07008 (2006);
M.E. Cates and S. Ramaswamy, cond-mat/0511260.

\bibitem{science} L. Berthier, G. Biroli, J.-P. Bouchaud, L. Cipelletti,  
D. El Masri, D. L'H{\^o}te, F. Ladieu, and M. Pierno, 
Science {\bf 310},  
1797 (2005). 

\bibitem{mct} W. G{\"o}tze, J. Phys. Cond. Matt. {\bf 11}, A1 (1999).

\bibitem{bethe}
M. Sellitto, G. Biroli, and C. Toninelli, 
Europhys. Lett. {\bf 69}, 496 (2005).

\bibitem{FP} S. Franz, G. Parisi, 
J. Phys.: Condens. Matter {\bf 12}, 6335 (2000). 

\bibitem{silvio2} 
S. Franz, C. Donati, G. Parisi, and S.~C. Glotzer,  
Phil. Mag. B, {\bf 79}, 1827,  
(1999).

\bibitem{glotzer} 
C. Bennemann, C. Donati, J. Baschnagel, S.~C. Glotzer, 
Nature {\bf 399}, 246 (1999). 

\bibitem{lacevic} 
N. La\v{c}evi\'{c}, F.~W. Starr, T.~B. Schroder, and S.~C. Glotzer,  
J.Chem.Phys. {\bf 119}, 7372 (2003). 

\bibitem{toni} C. Toninelli, M. Wyart, G. Biroli, L. Berthier, 
J.-P. Bouchaud, Phys. Rev. E {\bf 71}, 041505 (2005). 

\bibitem{mayer} P. Mayer, H. Bissig, L. Berthier, L. Cipelletti, 
J.P. Garrahan, P. Sollich, and V. Trappe,
Phys. Rev. Lett. {\bf 93}, 115701 (2004). 

\end{thebibliography}
\end{document}